\documentstyle[11pt,newpasp,twoside,epsf]{article}
\def\edcomment#1{\iffalse\marginpar{\raggedright\sl#1\/}\else\relax\fi}
\reversemarginpar

\begin{document}
\title{Clusters of Galaxies at High Redshift: The LMT/GTM Perspective.}
\author{Omar L\'opez-Cruz \& Enrique Gazta\~naga}
\affil{INAOE, Apdo. Postal 51 y 216, Tonantzintla, Puebla,
PUE 72000, Mexico}

\begin{abstract}
The lack of reliable cluster samples at intermediate and high redshift
has motivated many optical/infrared and X-ray cluster searches.  The
motivation of such searches is well justified: the abundance of
massive ($\approx 10^{15}\, {\rm M}_{\sun}$) collapsed regions gives a
direct constraint on $\Omega_{matter}$. Alternatively, the
Sunyaev-Zeldovich effect (SZE) offers a robust method for the
detection of clusters which is, almost, independent of redshift. In
this paper we comment on the possibility of mapping the SZE at high
resolution in the millimeter regime using the Large Millimeter
Telescope/Gran Telescopio Milim\'etrico (LMT/GTM) and the bolometer
array BOLOCAM.  The construction of the LMT/GTM facility is underway
atop Cerro La Negra (latitude=$18\deg59\arcmin$; height=4600 m) in
Mexico.
\end{abstract}
\keywords{Apj Keywords: <http://www.noao.edu/apj/keywords96.html>}
\section{Introduction}

One of the running themes of this conference is the hidden Universe
behind the Milky Way. We have seen throughout the conference how
successful some searches have been at detecting clusters of galaxies
in the Zone of Avoidance (e.g.~Ebeling et al., in these
proceedings).  Tantamount to a Zone of Avoidance the Universe at
high redshift remains unchartered. However, the millimeter regime
offers a new window to explore the high-$z$ regions and uncover its
large scale structure. The interaction of the cosmic microwave
background (CMB) and the hot plasma trapped in clusters gives a
distinctive signature in the resulting distorted spectrum that is very
different from other CMB distortions. This uniqueness of the resulting
spectrum gives a very efficient tool to detect clusters of galaxies at
any redshift. The LMT/GTM project with the bolometer array camera
BOLOCAM will capitalize on the ability to detect the SZE; hence, the
possibility to conduct large surveys will be feasible. One mode will
be to make pointed observations of nearby clusters of galaxies to make
high resolution maps of the SZE, these observations will help to
measure the baryonic content of clusters, H$_{0}$, and their peculiar
velocities through the detection kinematic SZE. In the second
mode of observation, large regions of the sky will be mapped in three
colors to search for clusters and superclusters at high-$z$. The
abundance of clusters of galaxies gives a direct constraint on
$\Omega_{matter}$. Below we outline the formalism for the treatment of
the SZE effect (\S2), review the LMT/GTM project (\S3), and finally we
describe how the LMT/GTM + BOLOCAM could be turned into a powerful
machine for cluster cosmology (\S4).

\section{The Sunyaev-Zeldovich Effect}

Clusters of galaxies are the largest bound structures in the Universe
(e.g. Abell 3627).  They are also reservoirs of large amounts of
baryons and dark matter ($\rm M_{cluster}\sim 10^{15}\,M_{\sun}$).
Trapped in the potential wells of clusters of galaxies lies a hot,
tenuous and fully ionized gas ($T_{e}=10^8~{\rm K},
n_{e}=10^{-3}\,{\rm cm}^{-3}$) that is the major component of the
intracluster medium (ICM) and the clusters' baryonic content
(e.g.~Sarazin 1988).  Shortly after the detection of the cosmic
microwave background (CMB; $T_{CMB}=2.728\pm0.002\,{\rm K}$), Sunyaev
\& Zeldovich showed that the CMB photons and the electrons in the ICM can
interact through the inverse Compton effect resulting in a scattering
process. This process preserves the number of CMB photons, but gives a
net energy gain to the CMB photons in the cluster's line of sight. Thus
the CMB is spectrally distorted. This process, also known as the
thermal SZE, was originally proposed as a major source of mm
radiation. However, only recently the advances in detector technology
have allowed us to unambiguously measure this effect and use it as a
cosmological tool (see reviews by Birkinshaw 1999 and Carlstrom et
al.~1999).

\begin{figure}[h]
\plotfiddle{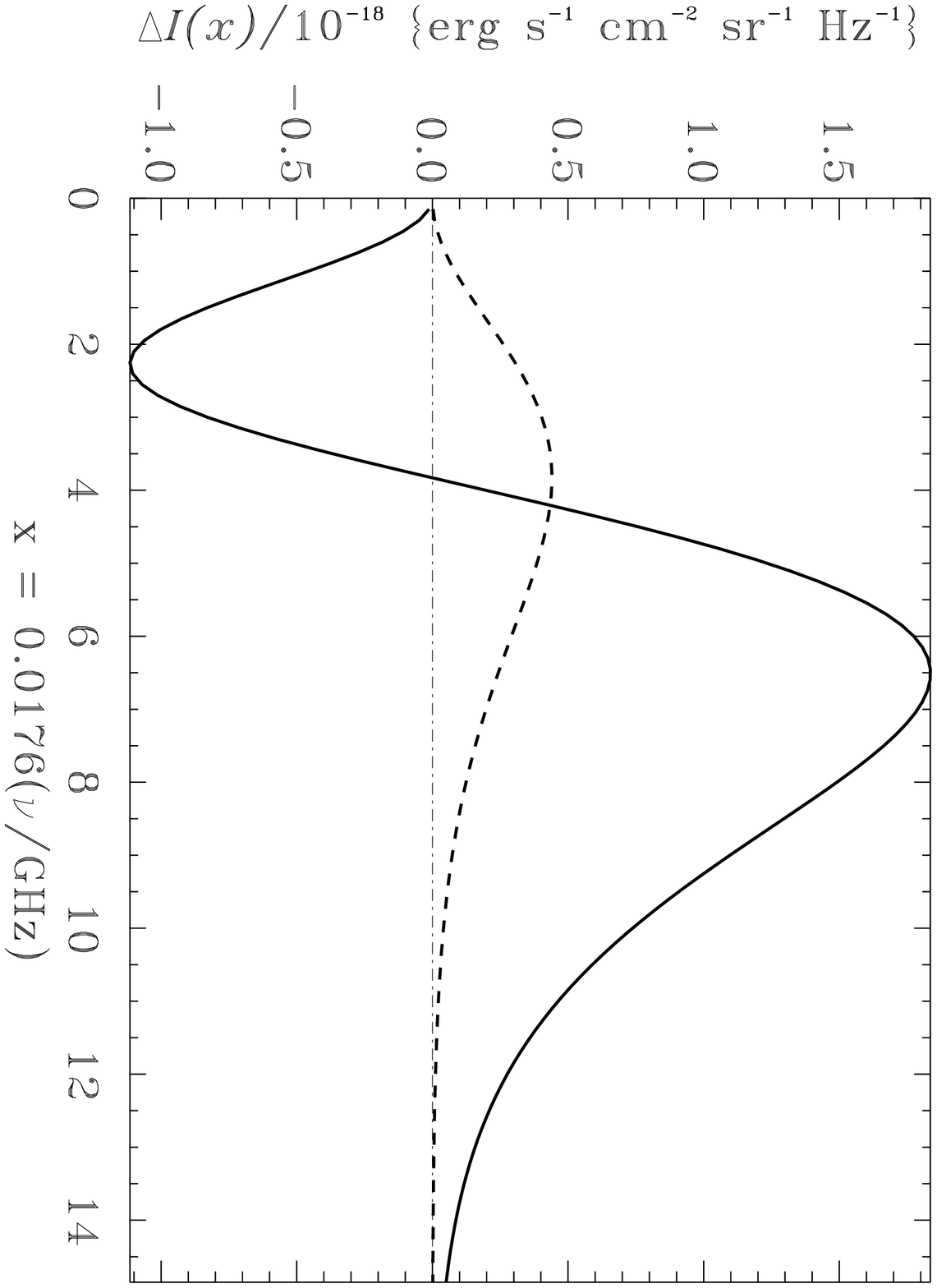}{6cm}{90}{45}{40}{170}{-25}
\caption{Spectral distortion of the CMB radiation due to the SZE. The thick
solid line represents thermal SZE ($y=10^{-4}$) and the dashed line is
the kinematic SZE ($\tau_{e}=10^{-2}$ and $v_{r}=-1000\rm\,
km\,s^{-1}$) for conditions found in massive clusters of galaxies.}
\end{figure}

The resulting spectrum of the CMB distortion due to inverse Compton
scattering in the non-relativistic regime at the limit
$k_{B}T_{e}>>h\nu$, where $h$ is the Planck constant, $\nu$ is the
frequency, and $k_{B}$ is the Boltzmann constant, is given by the
following equation (Zeldovich \& Sunyaev 1969):
\begin{equation}
g(x)= \frac{x^4e^x}{(e^x-1)^2}\left[\frac{x}{\tanh(x/2)}-4\right],
\end{equation}
where the
dimensionless frequency $x\equiv \frac{h\nu}{k_{B}T_{CMB}}$. The change of the CMB brightness $\Delta I_{T}(x)$, or 
temperature $\Delta T_{T}$, is
\begin{eqnarray}
\Delta I_{T}(x)&=&yI_{o}g(x),\\
\Delta T_{T}(x)&=&-yT_{CMB}\left[\frac{x}{\tanh(x/2)}-4\right],
\end{eqnarray} 
where $I_{o}\equiv\frac{2(k_{B}T_{CMB})^3}{(hc)^2}$. The amplitude of the effect is given by $y$.  This parameter is known
as the Comptonization parameter and is defined as follows:
\begin{equation}
y=\int\,\left(\frac{k_{B}T_{e}}{m_{e}c^2}\right)n_{e}\sigma_{T}dl
\end{equation}
where $m_{e}$ is the rest mass of the electron, $\sigma_{T}$ is the
Thompson cross section, and $n_{e}$ is the electron number density
that depends on physical conditions of the gas. In the case of the
{\em isothermal beta model} (Cavaliere \& Fusco-Femiano 1976) the
electron temperature is a constant and the electron density follows a
spherical distribution
\begin{equation}
n_{e}(r)= n_{e0}\left(1+\frac{r^2}{r_c^2}\right)^{-\frac{3}{2}\beta},
\end{equation}
where $n_{e0}$ is the central number density, $r_{c}$ is the ``core
radius'', and $\beta$ is the square of the ratio of the average galaxy
and gas particle speeds (Sarazin 1988). The condition of
isothermallity seems to hold in real clusters inside $\sim 30\%$ of
the virial radius (Irwin \& Bregman 2000).  Hence, the amplitude $y$
depends strongly on the cluster properties, i.e.~its temperature and
density structure; for massive clusters typical central values of
$y=10^{-4}$ are found. Nonetheless, the shape of the spectrum is
independent of the physical conditions of the clusters, but vary with
$T_{e}$ in the more complicated relativistic case (Sazonov \& Sunyaev
1998). The critical points of $\Delta I_{T}(x)$ are: a maximum at
$x_{max}=6.51\leadsto\lambda_{max}=0.8\,{\rm mm}$, i.e.~maximum
emission, a root at $x_{zero}=3.83\leadsto\lambda_{zero}=1.4\,{\rm
mm}$, i.e.~a null, and a minimum at
$x_{min}=2.26\leadsto\lambda_{min}=2.3\,{\rm mm}$, and a minimum at
$x_{min}=2.26\leadsto\lambda_{min}=2.3\,{\rm mm}$, i.e.~maximum
decrement. Figure 1 shows the behavior of the thermal SZE for a massive
cluster with $y=10^{-4}$.

The kinematic component of the SZE is due to the bulk motion of the
cluster, and its ICM, relative to the CMB rest frame. The electrons
will be scattered by the ICM electrons as above. However, if the
cluster has a proper motion, the scattered photons will be Doppler
shifted by an amount that depends on the angle of their scattering
relative to the bulk velocity. This effect is equivalent to a change
in the CMB temperature ($\Delta T_{K}$). Its spectrum is given by:
\begin{equation}
h(x)=\frac{x^{4}e^{x}}{(e^{x}-1)^2}.
\end{equation}
The change in the CMB brightness $\Delta I_{K}(x)$ due to the
kinematic SZE is
\begin{eqnarray}
\Delta I_{K}(x)&=&-I_{o}h(x)\int\frac{v_{r}}{c}n_{e}\sigma_{T}dl=-
\frac{v_{r}}{c}\tau_{e}I_{o}h(x)\\
\Delta T_{K}(x)&=&-\frac{v_{r}}{c}\tau_{e}\frac{x^{2}e^{x}}{(e^{x}-1)^2}
\end{eqnarray}
where $v_{r}= \mbox{\boldmath $\vec{v}_{p}\cdot\hat{r}$}$ is the
component of the bulk velocity along the line of sight towards the
observer and $\tau_{e}=\int n_{e}\sigma_{T}dr$. The kinematic SZE is
much smaller than the thermal SZE given the known properties of
clusters. At long wavelengths the kinematic SZE gets swamped by the
thermal SZE. However around $\lambda= 1$ mm the kinematic SZE reaches
its maximum change in brightness while the thermal effect goes to
zero. Figure 1 shows the kinematic SZE for a cluster with
$\tau_{e}=0.01$ and a projected peculiar velocity $v_{r}=-1000\, \rm
km\,s^{-1}$. We remark, therefore, on the relevance of submm-mm
observations for the detection of both components of the SZE.

\section{The LMT/GTM Project}

\begin{figure}[h]
\plotfiddle{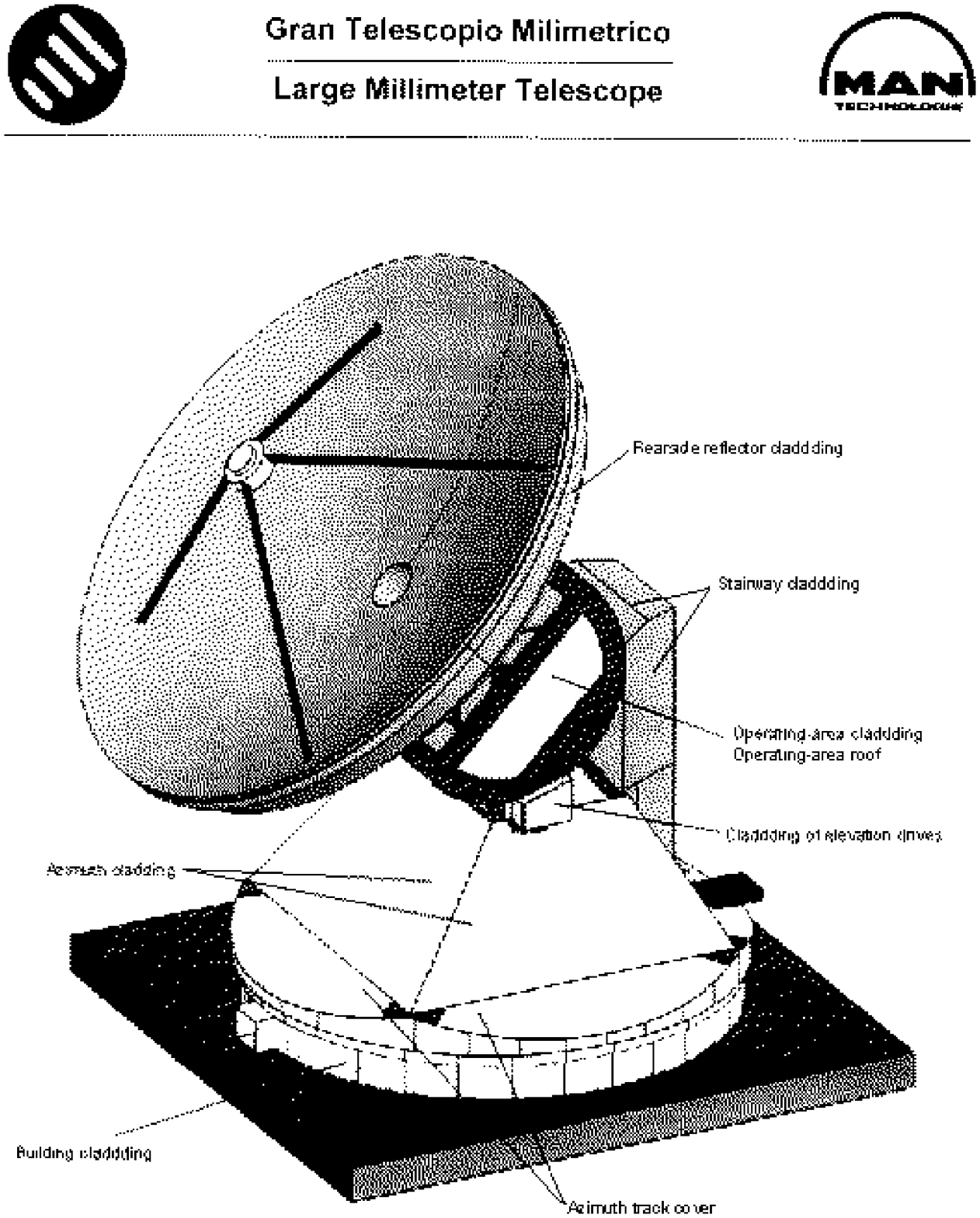}{10cm}{0}{38}{38}{-100}{-20}
\caption{This is an artist's view of the Large Millimeter
Telescope/Gran Telescopio Milim\'etrico (GTM/LMT). With an aperture of
50 m the GTM/LMT will be the largest telescope of its kind. The
technical demands for the antenna to operate efficiently in the mm
regime (0.85-3.4 mm) are very high. The pointing and tracking accuracy
has to be $\sim 0.7\arcsec$ under wind conditions of $< 6$\,m/s (50\%
of the time at Cerro La Negra). The projected final surface rms
accuracy $\sigma=70\,\micron$ will result in the aperture efficiencies
$\eta_{\lambda=1\,{\rm mm}}=46\,\%$ and $\eta_{\lambda=3\,{\rm
mm}}=75\,\%$.}
\end{figure}

The LMT/GTM is a joint collaboration between the United States of
America and Mexico.  This is the largest scientific collaboration
ever tried out between these two countries. The project contemplates
the design, construction, and operation of a 50 m antenna that will be
sensitive in the millimeter wavelength regime ($\lambda=0.85-3.4\,\rm
mm$). Figure 2 depicts an artist's view of such an antenna. The main
institutions that are involved in the project are the University of
Massachusetts (UMass) and the Instituto Nacional de Astrof{\'\i}sica,
Optica y Electr\'onica (INAOE). For further details visit the web site
$<$http://www.lmtgtm.org$>$. First light is expected in the year 2002.

\subsection{The Telescope Site}

Millimeter astronomy poses more challenges than radio astronomy. The
atmosphere is more opaque in the mm regime and is highly variable. An
extensive search was conducted in many mountains in Mexico searching
for places with high transparency and low water vapor column densities.
The selected site was Cerro La Negra, a 4640\,m volcanic mountain in
the State of Puebla (lat. $18\deg\,59^{'}$). This site is one of the
highest peaks in the Mexican Altiplano, located south of both Cofre de
Perote and Pico de Orizaba. Cerro La Negra is an excellent site for mm
observations. About $>65\%$ of the time between October and
May it has a $\rm \tau_{210\,GHz} <0.2$ with a 1$^{st}$ quartile
$< 0.08$.  850$\micron$ observations will also be possible during the
best conditions ($\leq 10\%$) when $\rm \tau_{340\,GHz}$ is $<
0.2$. Additionally, during the summer months Cerro La Negra remains an
excellent 3\,mm site.

\subsection{The Antenna}

Very stringent antenna requirements need to be met to efficiently
explore the mm window and take full advantage of Cerro La Negra
conditions. They also pose a strong technological challenge because a
50\,m aperture antenna for the millimeter has never been built before.
For instance, the required pointing and tracking accuracy has to be
$\sim 0.7$\,arcsecs under wind conditions of~$< 6$\,m/s (50\% of the
time at Cerro La Negra). The demands on the quality of the surface are
also high:  for a projected initial rms surface accuracy $\sigma=100\,
\micron$, the resulting aperture efficiency is $\eta_{\lambda=1\,{\rm
mm}}=21\%$ at $\lambda=1\,$mm. However the eventual goal is to reach
$\sigma=70\,\micron$ that will give a good aperture efficiency
$\eta_{\lambda=1\,{\rm mm}}=46\,\%$ and an optimal
$\eta_{\lambda=3\,{\rm mm}}=75\,\%$.

\subsection{BOLOCAM}

Analogous to the evolution in detector technology that was seen in the
near infrared during last decade, arrays with increasing numbers of
elements have made their debut into submm-mm astronomy. We can
mention for example, the six element SuZIE (Holzapfel et al.~1997)
at the Caltech Submillimeter Observatory (CSO), the 37/91 element
SCUBA (Holland et al.~1999) at the James Clerk Maxwell Telescope
(JCMT), and the six element Diabolo (Benoit et al.~2000) experiment at
Institute de Radioastronomie Millim\'etrique (IRAM) 30\,m, among
others. BOLOCAM is designed around a monolithic wafer containing 151
silicon nitride micromesh bolometers arranged in a hexagonal pattern
with a center-to-center spacing of 5\,mm (Glenn et al.~1998). The
detectors provide photon noise limited performance at $\lambda=1\,$mm
at a base temperature of 30 mK. The feed horns and cavities are
machined into single pieces of aluminum and invar. The camera design
and optics are presented in Fig.~3.  The first prototype was
completed early this year and has been tested at CSO. The second one
will be commissioned at the LMT/GTM. The LMT/GTM + BOLOCAM will have a
field of view of 2$\arcmin$ and a resolution $\rm
\theta_{beam}^{\lambda=1.4mm}\approx 8\,\arcsec$ and an $NEFD\footnote{ Noise-Equivalent-Flux-Density}\,\sim\,2.8\,
\rm mJy/Hz^{1/2}$ at 1.4 mm (220 GHz), while at 1.1 mm (271 GHz) the
resolution $\rm \theta_{beam}^{\lambda=1.1 mm}\approx 7\,\arcsec$. The
projected mapping speed at 1.1\,mm will be about $2\,
\rm deg^2\,hr^{-1}$. BOLOCAM will have three bandpasses centered on
$\lambda= \rm 1.1,1.4, \,and \,2.1\, mm$; however, for technical
reasons only one passband can be used for each observing run (Glenn et
al. 1998). The choice of the $\lambda= \rm 1.1 \,and \,2.1 mm$
wavebands maximizes the ratio of the SZE thermal component signal to
the combination of atmosphere and detector noise. The $\lambda=1.4$
was chosen to detect the strongest component of the kinetic SZE while
minimizing the component of the much stronger thermal SZE (see Holzapfel
et al. 1997, for further details).

\begin{figure}
\plotfiddle{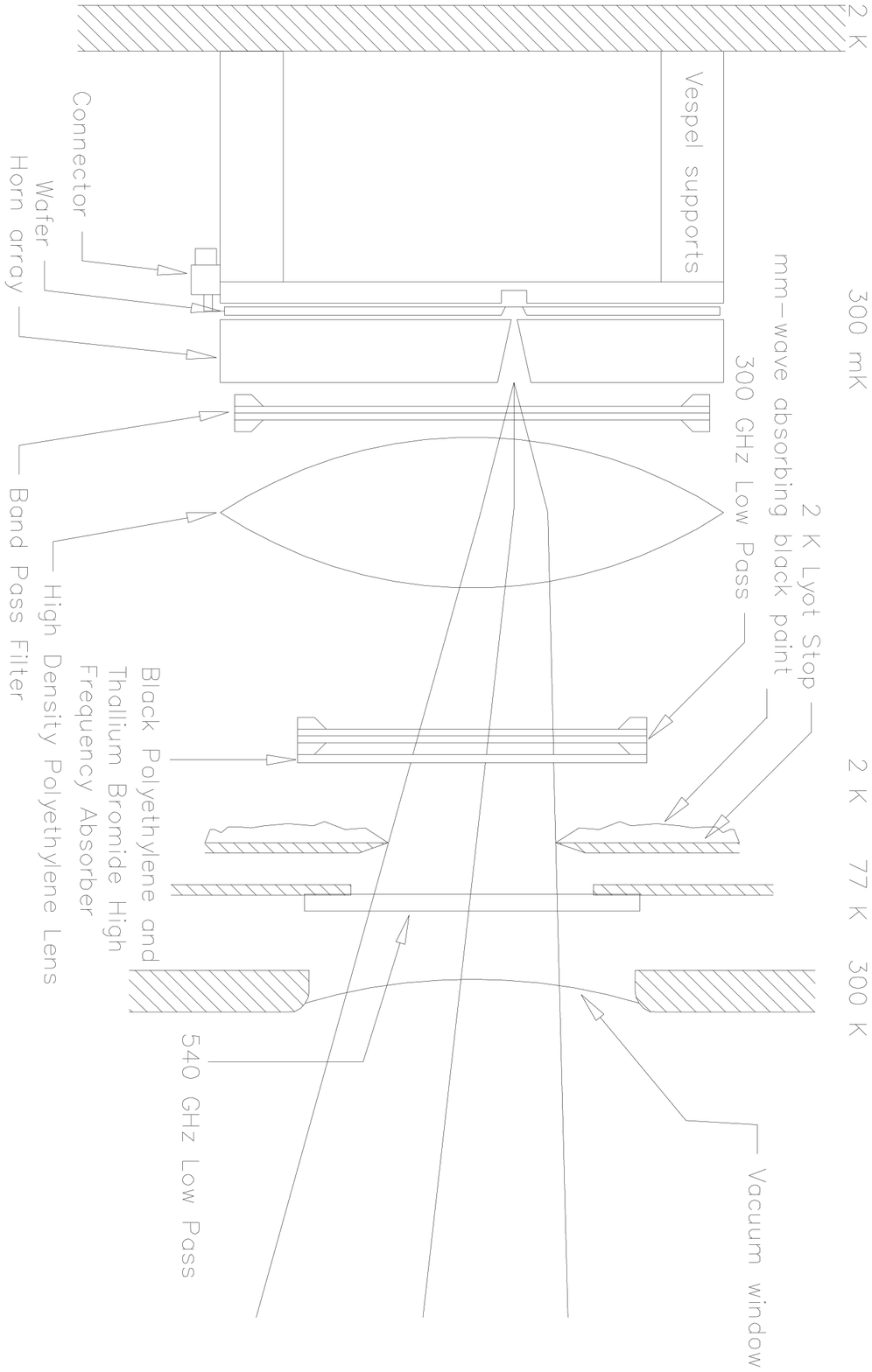}{6cm}{90}{45}{45}{170}{-45}
\caption{BOLOCAM camera design and cold optics.}
\end{figure}

\section{Search for Clusters of Galaxies at High Redshift}

Clusters of galaxies are important probes of the large scale structure
of the Universe. As they originate from the largest high density
enhancements in the Cosmic Web (Bond, Kofman, and Pogosyan 1996, BKP):
their abundance sets direct constraints on $\Omega_{matter}$ 
(e.g.~Barbosa et al.~1996, Holder et al.~2000, Haiman et al.~2000) and the
amplitude of the primordial fluctuations ($\sigma_8$) (BKP,
Bond et al.~1998, Haiman et al.~2000). Therefore, is very important
from the observational point of view to arrive at the cleanest sample
within any surveyed area. The thermal SZE, being independent of
redshift and having a characteristic spectrum, provides an ideal tool
for cluster detection.

Many cluster searches are currently underway either in the optical
(e.g.~Gladders \& Yee 2000) or in the X-ray (e.g.~Ebeling et al.~2000).
Among many other biases that those searches might have, the
most important one is that the detectability has a strong dependence
with redshift ($\propto (1+z)^{-4}$) while the mass sensitivity of the
thermal SZE has a weaker dependence with redshift. The flux density
due to the thermal SZE is the integral of the spectral distortion
integrated over the solid angle subtended by the cluster:

\begin{equation}
S_{\nu}=g(x)D_{A}^{-2}\int\,\left(\frac{k_{B}T_{e}}{m_{e}c^2}\right)n_{e}\sigma_{T}dV;
\end{equation}
where the integral is over the cluster volume and
$D_{A}=\frac{2c}{H_{o}\Omega_{o}^2}
\frac{\Omega_{o}z+(\Omega_{o}-2)(\sqrt{1+\Omega_{o}z}-1)}{(1+z)^2}$
is the angular diameter distance.  Hence the flux depends on the
total number of electrons at temperature $T_{e}$ and the angular
distance only. Simple scaling laws show that
$f_{sze}\sim(1+z)^{3/2}f_{x}$, where $f_{sze}$ and $f_x$ are the
fluxes due to the SZE and X-ray respectively. Hence, the SZE probes
deeper in redshift than the X-rays.

\begin{figure}[h]
\plotfiddle{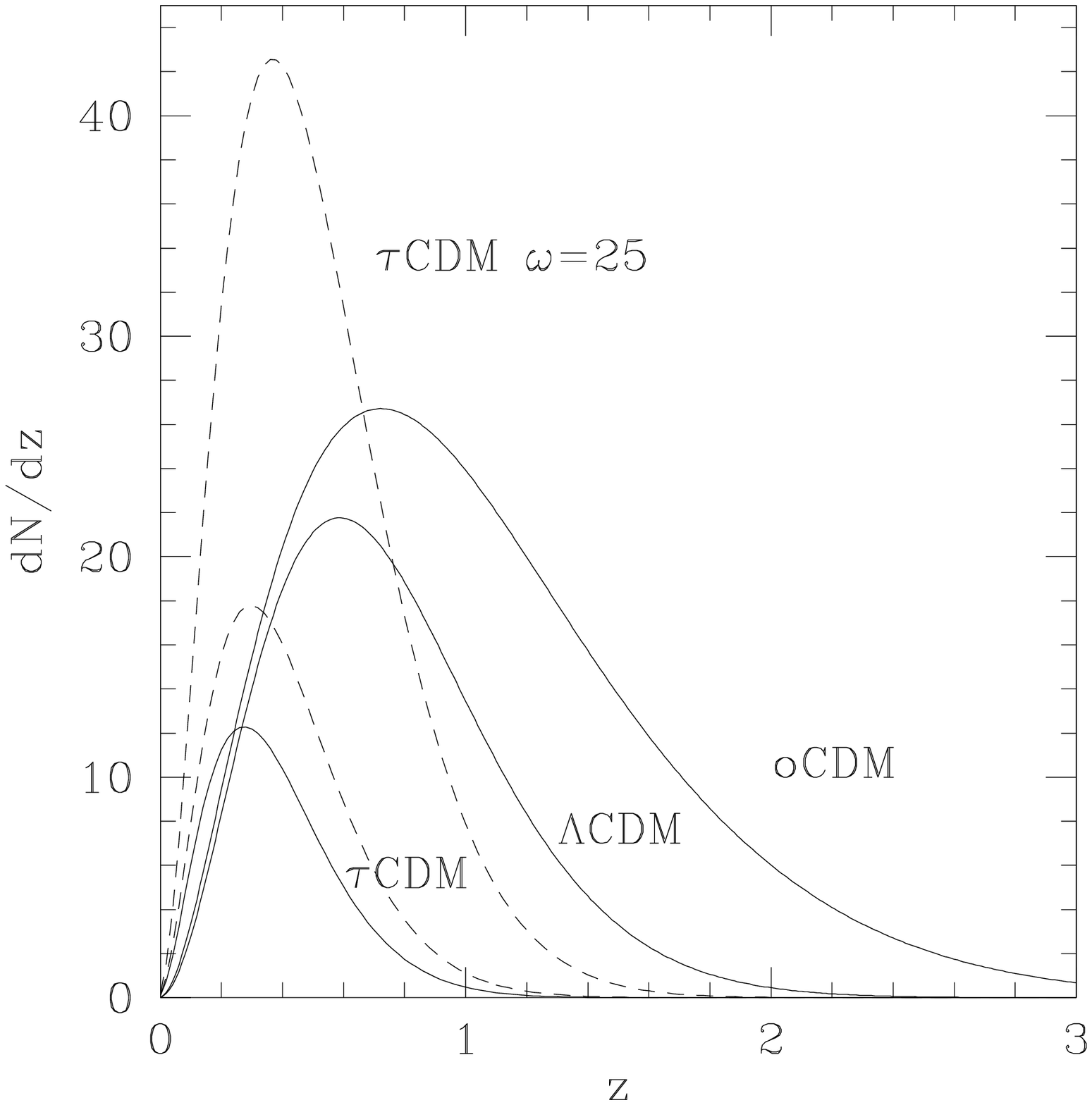}{5cm}{0}{55}{32}{-180}{-65}
\caption{ The solid lines represent the expected differential 
count  distribution per
square degree of massive clusters ($\rm
M > M_{th} \times 10^{14}\,h^{-1}_{50}{M_{\sun}} $)
for three cosmologies oCDM
($M_{th}=1.9 $,
$\Omega_{m}=0.3,\Omega_{\Lambda}=0, h=0.65, \Gamma=0.25,
\sigma_{8}=1.0$); $\Lambda$CDM ($M_{th}=2.2$,
$\Omega_{m}=0.3,\Omega_{\Lambda}=0.7,
h=0.65, \Gamma=0.25, \sigma_{8}=1.0$) and $\tau$CDM
($M_{th}=1.3$,
$\Omega_{m}=1.0;\Omega_{\Lambda}=0, h=0.5, \Gamma=0.25,
\sigma_{8}=0.56$) derived using the Press-Schechter prescription (see
Holder et al.~2000). The lower (upper) dashed lines 
correspond to a Brans-Dicke Cosmology with $\omega=100$  ($\omega=25$)
normalized to COBE with the $\tau$CDM model
(see details in Gazta\~naga \& Lobo 2000).}
\end{figure}

During routine operation LMT/GTM + BOLOCAM will be able to map massive
clusters ($y=10^{-4}$) at high $z$ with a resolution of $7\arcsec$ at
1.1 mm in a few hours.  Assuming that a $z=3$ massive cluster has 
$r_{c}\sim 300$ kpc then $S_{\nu}=0.1-0.2
\rm mJy/beam$ (H$_{0}=50\,h^{-1}{50}$ km\,s$^{-1}$,Mpc$^{-1}$;
$\Omega_{o}=0.2$). Hence, it will be possible to map the entire
cluster and sorrounding sky over an area of $3-4\,\rm arcmin^2$ down
to the $3\sigma$ sensitivity level in about 2 hrs. A 200 hrs survey,
after commissioning, could cover a 10 ${\rm deg^2}$ region to a flux
limit of $~1$ mJy. Depending on the cosmology this survey could yield
about 50-500 massive clusters (M $> 2\times 10^{14}\,M_{\sun}$).  In
Fig.~4 remarkable differences in the cluster counts can be seen
between different, standard, cosmological models at $z>1$ .Deviations from
General Relativity can be noticed even at lower redshifts when models are
normalized to CMB fluctuations (Gazta\~naga
\& Lobo 2000).  Observations at 1.1 mm and 2.4 mm could separate the
fluctuations due to clusters from the other cosmological
fluctuations. This way it would be possible to gather the largest
reliable mass selected high-$z$ cluster samples.

Follow up observations with large infrared telescopes and the new X-ray
missions will be needed to determine the mass and redshift
distribution of the detected clusters. These observations will set
strong constraints on the matter content of the Universe. At the same
time we will also be exploring the large scale distribution of the
Universe, gas distribution, and galaxy evolution at unprecedented
high-$z$.

\acknowledgements
OLC is grateful to the organizing committee for support to participate
in this very interesting conference.  The research of OLC is supported
by CONACyT (J-32098-E) and ANUIES-ECOS. OLC wishes to thank Gil
Holder, Steve Myers and Chris Carilli for enlighting discussions. The
hospitality of the PUC and
Observatoire de Marseille is gratefully acknowledged.


\begin{references}
\reference
Barbosa, D., Bartlett, J.G., Blanchard, A., \& Oukbir, J. 1996, \aap, 314, 13

\reference
Benoit, A., Zagury, F., Coron, N., et al.~2000, \aaps, 141, 523 

\reference
Birkinshaw, M. 1999, Physics Reports, 310, num.~2-3, 98


\reference
Bond, J.R., Kofman, L., \& Pogosyan, D. 1996, Nature, 380, 603 

\reference
Bond, J.R. et al.~1998, in  Wide Field Surveys in Cosmology, eds. S. Colombi,
Y. Mellier \& B. Raban, (Gif-sur-Yvette:  Editions Frontiers), 17 

\reference
Carlstrom, J.E., Joy, M.K., Grego, L., et al. ~1999, astro-ph/9905255

\reference
Cavaliere, A. \& Fusco-Femiano, R. 1976, \aap, 49, 137

\reference
Ebeling, H., Edge A., \&  Henry, J.P. 2000, astro-ph/0001320

\reference
Gazta\~naga, E.,  Lobo, A. 2000, astro-ph/0003129
 
\reference
Gladders M.D. \& Yee H.K.C. 2000, astro-ph/0004092 

\reference
Glenn, J., Bock, J.J., Chattopadhyay, G., et al.~1998, Proc. SPIE, 3357,326

\reference
Haiman, Z., Mohr, J.J., \& Holder, G.P. 2000, astro-ph/0002336 

\reference
Holder, G.P., Mohr, J.J., Carlstrom, J.E., et al. ~2000,
Apj in press.

\reference
Holland, W.S., Robson, E.I., Gear, W.K., et al.~ 1999, \mnras, 303,659

\reference
Holzapfel, W. L., Wilbanks, T. M., Ade, P. A.,  et al.~1997, \apj, 479, 17

\reference
Irwin, J.A. \& Bregman, J. 2000, astro/ph0003123

\reference
Sarazin, C. L. 1988,  X-ray Emission from Clusters of Galaxies. 
(Cambridge:  Cambridge University Press)

\reference
Sazonov, S.Y. \& Sunyaev, R.A. 1998, \apj, 508, 1

\reference
Zeldovich, Ya. B. \& Sunyaev, R.A.  1969, \apss, 4, 301 

\end{references}
\end{document}